\documentclass[journal,subeqn,caption]{IEEEtran}
\usepackage[utf8]{inputenc}
\usepackage{amsmath}
\usepackage{bm}
\usepackage{amsfonts}
\usepackage{amssymb}
\usepackage[pdftex]{graphicx}
\usepackage{forest}
\usepackage{mathrsfs}
\usepackage{mathtools}
\usepackage[ruled]{algorithm2e}
\usepackage{cite}
\usepackage{multirow}
\usepackage{diagbox}

\newcommand{\E}[1]{\mathbb{E}\big\{ #1 \big\}}
\newcommand{\Prob}[1]{\mathbb{P}( #1 )}
\newcommand{\Pbig}[1]{\mathbb{P}\big( #1 \big)}

\newcommand{\AmbSet}{\hat{\mathcal{P}}_K}
\newcommand{\EmpDist}{\hat{\mathbb{P}}_K}

\usepackage{amsmath,amsfonts,amsthm}

\theoremstyle{definition} 

\theoremstyle{plain} 
\newtheorem{theorem}{Theorem}

\newtheorem{corollary}{Corollary}

\theoremstyle{remark} 

\newcommand{\norm}[1]{\left\lVert#1\right\rVert} 

\DeclareMathOperator{\CVaR}{CVaR}
\DeclareMathOperator{\VaR}{VaR}

\def\myproofstart{\noindent\textit{Proof. }}
\def\myproofend{\hspace*{\fill} $\square$ \vspace{0.3\baselineskip}}
\usepackage{placeins}
\newcommand{\subparagraph}{}
\usepackage{titlesec}

\titlespacing*{\subsection}{0pt}{4pt}{1pt}

\title{Co-Control of VaR and CVaR for \\ Data-Driven Stochastic Demand Response Auction}
\author{Matt Roveto, Robert Mieth, Yury Dvorkin}

\begin{document}
\maketitle

\begin{abstract}
The ability to make optimal decisions under uncertainty remains important across a variety of disciplines from portfolio management to power engineering.
This generally implies applying some safety margins on uncertain parameters that may only be observable through a finite set of historical samples.
Nevertheless, the optimized decisions must be resilient to all probable outcomes, while ideally providing some measure of severity of any potential violations in the less probable outcomes.
It is known that the conditional value-at-risk (CVaR) can be used to quantify risk in an optimization task, though may also impose overly conservative margins. 
Therefore, this paper develops a means of co-controlling the value-at-risk (VaR) level associated with the CVaR to guarantee resilience in probable cases while providing a measure of the average violation in less probable cases.
To further combat uncertainty, the CVaR and VaR co-control is extended in a distributionally robust manner using the Wasserstein metric to establish an ambiguity set constructed from finite samples, which   is guaranteed to contain the true distribution with a certain confidence.
\end{abstract}

\section{Introduction}\label{sec:intro}
In power grid applications, optimal decision making under uncertainty requires models and solution methods that appropriately internalize the risk associated with stochastic parameters, e.g. scenario-wise Stochastic Programming (SSP) \cite{pritchard2010single, papavasiliou2011reserve} and Chance-Constrained Optimization (CC) \cite{zheng2013decomposition, bienstock2014chance}.
Resulting solutions have been shown to outperform those of deterministic methods, e.g by reducing overly conservative, exogenous safety margins while maintaining constraint satisfaction with a high  probability and remaining tractable, \cite{bienstock2014chance}.

Additionally, distributionally robust optimization (DRO) internalizes uncertainty on the distribution of the uncertain stochastic parameters by solving over an ambiguity set characterized by limited information on the unknown distribution.
Thus DRO solutions are more resilient to errors that may arise from incorrect assumptions on the underlying distribution, and avoid solution conservatism arising from considering the worst case of all possible outcomes.
Depending on the constructed ambiguity set, various tractable and scalable methods to solve DRO problems have been proposed, e.g. \cite{ben2009robust,lubin2015robust,zhao2018data}.
Although the most popular choice of ambiguity sets are generated via distributions with a given empirical mean and variance \cite{delage2010distributionally,erdougan2006ambiguous,mieth2018data}, such moment-based sets ignore a significant amount of available information on the underlying distribution.
Metric-based ambiguity sets, on the other hand, are characterized as a set in the space of probability distributions by using a ``distance'' measure (e.g. the Prohorov metric, Kullback-Leibler divergence, or Wasserstein metric), and offer an appealing replacement to moment-based ambiguity sets, \cite{esfahani2018data}.

In most stochastic optimization methods decision-dependent, uncertain (random), outcomes are replaced with a deterministic equivalent that evaluates the underlying stochastic processes using a measure of risk.
Most prominently, the expectation operator replaces the uncertain outcome of a decision (e.g. future cost) with the average outcome of infinite repetitions of the decision with respect to some underlying probability distribution.
Alternatively, the value-at-risk (VaR) evaluates the uncertain outcome in terms of the probability that it will not exceed a predefined threshold.
Leveraging this property, the VaR also allows enforcing probabilistic (chance) constraints of the form $\mathbb{P}(\bm{X} \le b) \ge \alpha$ by requiring $\VaR_\alpha(\bm{X}) \le b$. 
Among large-scale, real-life applications, tractable closed form expressions exist for example in the context of optimal power flow (OPF) formulations, \cite{bienstock2014chance}, and thus have become a popular tool in power system analyses due to the close relation of such chance constraints with existing reliability measures.

\begin{figure}[!t]
    \centering
    \includegraphics[width=\columnwidth]{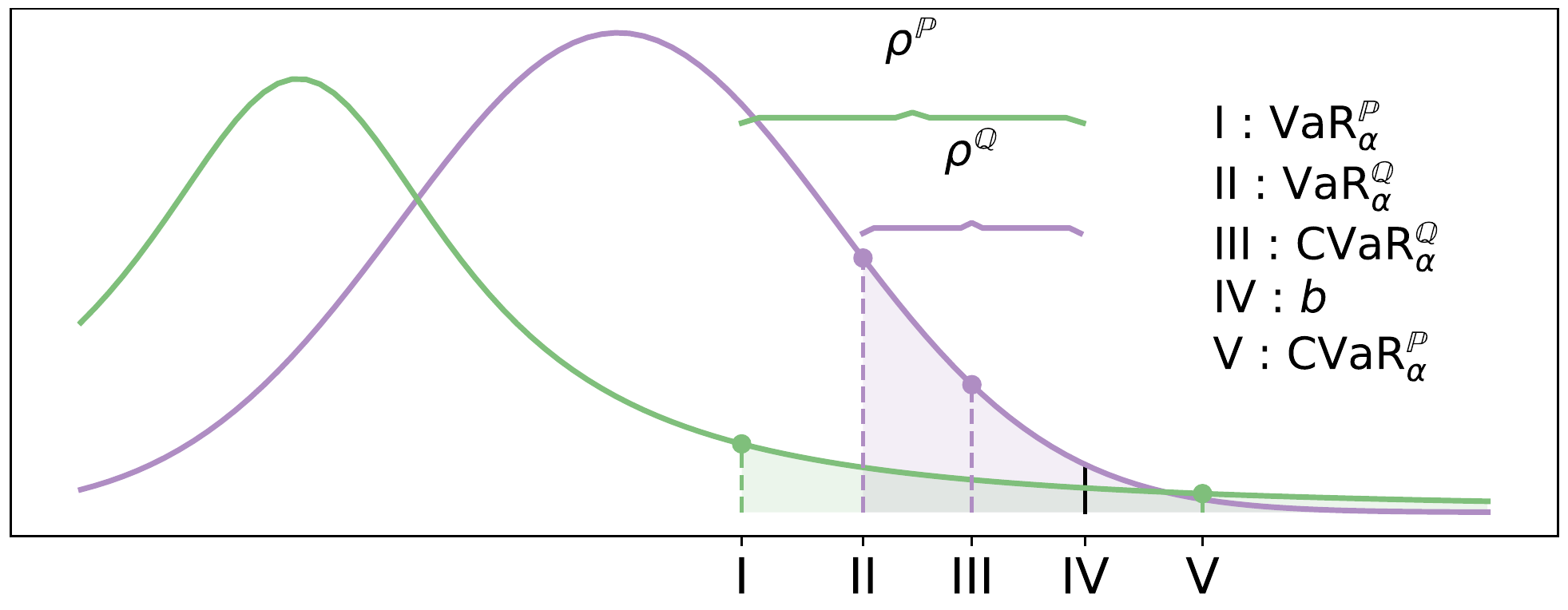}
    \vspace{-7mm}
    \caption{Probability density of a uncertain loss function governed by two different underlying probability distributions, in purple ($\mathbb{Q}$) and green ($\mathbb{P}$), and their respective $\VaR_\alpha$ and $\CVaR_\alpha$. 
    The shaded area represents the probabilty of events beyond the $\VaR_\alpha$.
    The acceptability threshold is denoted by $b$ and the distance between $b$ and the $\VaR$ is denoted by $\rho$. 
    While the $\VaR_\alpha$ of both beliefs lies below $b$, only the $\CVaR_\alpha^{\mathbb{Q}}$ of the purple distribution is below this bound. As such, if the true distribution follows the green curve, the losses may be much worse than expected, motivating the use of a distributionally robust approach incorporating the co-control of VaR and CVaR. \label{fig:VaRCVaRGraphic}}
\end{figure}

However, if the underlying probability distribution has to be inferred via sampling or scenarios the VaR has tractability issues, i.e. it obstructs attaining a global optimal solution due to nonconvexity and nondifferentiability, \cite{lim2010portfolio,rockafellar2007coherent}.
Alternatively, mostly enabled by the work of Rockafellar \textit{et al.}, \cite{rockafellar2000optimization,rockafellar2002conditional,rockafellar2007coherent}, the conditional value-at-risk (CVaR) may be calculated using convex optimization techniques and so is more practical to use.
Notably, optimizing the CVaR implies an optimization of the VaR, i.e. as $\VaR_\alpha(\bm{X})$ determines a upper bound on $\bm{X}$ with probability $\alpha$, $\CVaR_\alpha(\bm{X})$ is the expected value of $\bm{X}$ under the condition that $\bm{X}\ge\VaR_\alpha$ \cite{rockafellar2002conditional, rockafellar2007coherent}, as shown for two different distributions in Fig.~\ref{fig:VaRCVaRGraphic}.
This property can be leveraged as a means of providing a convex approximate upper bound on chance constraints, e.g. as shown in Dall'Anese \textit{et al}. \cite{dall2017chance} for a distributed chance-constrained OPF.

The $\VaR$ and $\CVaR$, however, may provide more than a bound on certain constraints, and offer an additional tool to handling uncertainty in the system.
Naively enforcing that $\CVaR_\alpha(\bm{X}) < b$, may lead to an overly conservative solution since $\VaR_\alpha \le \CVaR_\alpha$, \cite{rockafellar2002conditional, rockafellar2007coherent}, as shown in Fig.~\ref{fig:VaRCVaRGraphic}.
Thus, if a decision maker only requires constraints hold with $\alpha$ probability, a feasible solution of the objective may exist in which $\VaR_\alpha(\bm{X}) = b$ that improves upon the solution in which $\VaR_\alpha(\bm{X}) < b$, i.e. one in which $\rho$, the gap between $\VaR_\alpha(\bm{X})$ and $b$, is equal to zero (see Fig.~\ref{fig:VaRCVaRGraphic}).
This still enforces the desired chance constraints without accruing extra costs of added security.
As such, we would like to control the VaR from being lower than $b$, while minimizing the CVaR, i.e. co-controlling the VaR and CVaR with the superior mathematical properties of the CVaR.

This paper aims to synthesize the most desirable properties of VaR and CVaR by presenting a model that endogenizes considerations of risk and cost by co-controlling both measures.
In doing so, the solution will provide insights of how severely constraints may be violated via constrained minimization of the CVaR, while ensuring that the corresponding VaR does not produce an overly conservative solution.
By ensuring the VaR does not exceed some critical value of the stochastic constraint, this method will also simultaneously guarantee that the corresponding chance constraint holds with a desired probability.
Lastly, we introduce a DRO extension that hedges against over-specifying the distribution while using the smallest possible ambiguity set containing the true distribution with near certainty by means of the Wasserstein metric which is guaranteed to contain the true distribution under certain conditions
\cite{bolley2007quantitative, fournier2015rate}.
Using results of \cite{esfahani2018data, duan2018distributionally} the model scales well, and due to the convexity of CVaR calculations, preserves the convexity of problems allowing for solutions with off-the-shelf solvers.

We demonstrate the usefulness of this model through a stochastic reverse auction to procure demand response (DR) in a power system.
Here, one must ensure they acquire the minimum target amount of DR lest they receive a penalty proportional to the amount by which they have fallen short of the target, but must also do so at the lowest cost.
This is a particularly challenging task as they must procure the DR from a pool of uncertain bids.

\section{Data-Driven Co-Control of VaR and CVaR}\label{sec:CoControl}
\subsection{Preliminaries}\label{sec:Prelim}
Let us first define the $\alpha$-Value-at-Risk ($\VaR_\alpha$) of a a random vector $\bm{X}$ as the smallest possible value for which the probability that $\bm{X}$ is less than or equal to that value is at least $\alpha$, i.e.
\begin{align}
    \text{VaR}_\alpha(\bm{X}) \coloneqq \min\{x | \Prob{\bm{X} \le x} \ge \alpha\},\label{VaRDef}
\end{align}
where $\mathbb{P}$ is the probability measure of ${\bm{X}}$.
Notice then that by \eqref{VaRDef}, when $\mathbb{P}$ is continuous and strictly increasing, the VaR$_\alpha$ is the unique value satisfying $\Prob{\bm{X} \le x} = \alpha$, \cite{rockafellar2002conditional}.
We then define the CVaR$_\alpha$ of $\bm{X}$ as the mean of the $\alpha$-tail distribution of $\bm{X}$ as in \cite{rockafellar2002conditional}, i.e.
\begin{align}
    \CVaR_\alpha(\bm{X}) \coloneqq \E{\bm{X} | \bm{X} > \VaR_\alpha} .\label{CVaRDef}
\end{align}

Naturally then, we have that $\CVaR_\alpha(\bm{X}) \ge$ VaR$_{\alpha}(\bm{X})$,
and that $\Prob{\bm{X} \le \text{VaR}_{\alpha}} \ge \alpha$ from the definition of the $\VaR_\alpha$.
Following \cite{rockafellar2007coherent}, in order to practically calculate the $\CVaR_\alpha$ and $\VaR_\alpha$, we use the following minimization rule in which
\begin{align}
    \CVaR_\alpha(\bm{X}) = \min_{z \in \mathbb{R}}\{z + \frac{1}{1 - \alpha}\E{[\bm{X} - z]^+} \} \label{CVaRMinimization}
\end{align}
where $[t]^+ \coloneqq \max\{0,t\}$, and thus \eqref{CVaRMinimization} can be rewritten in terms of a piece-wise linear loss function
\begin{align}
    \CVaR_\alpha(\bm{X})& = \min_{z \in \mathbb{R}} \E{\max_{j\le J}l_{j}(\bm{X})}\label{SpecialFuncReform}
\end{align}
where $l_{j}(\bm{X}) \coloneqq a_{j}\bm{X} + b_{j}z$ for $J = 2$, $a_{1} = 0, b_{1} = 1, a_{2} = 1/(1 - \alpha), b_{2} = 1 - 1/(1 - \alpha)$, which is finite and convex (hence continuous) as a function of $z$ by \cite[Theorem 10]{rockafellar2002conditional}.
Note that while minimization of \eqref{SpecialFuncReform} yields the $\CVaR_\alpha$, the corresponding optimizer $z^*$ is equal to $\VaR(\bm{X})$, \cite{rockafellar2000optimization}.

Consider now a more general case in which the random variable is also dependent on some other deterministic decision variable $y$, i.e. $\bm{X}(y)$.
We are no longer interested in just the calculation of $\CVaR_\alpha$, but would like to find such $y$ that minimizes our $\CVaR_\alpha$.
To do so, we note that function $\E{\max_{j\le J}l_{j}(\bm{X})}$ is jointly convex over both $z$ and any convex functions of $\bm{X}$, such that by \cite[Theorem 14]{rockafellar2002conditional}, minimizing $\CVaR_\alpha(\bm{X}(y))$
with respect to $y \in Y$ is equivalent to minimizing $\E{\max_{j\le J}l_{j}(\bm{X}(y))}$ over all $\big(\bm{X}(y), z\big) \in Y \times \mathbb{R},$ such that
\begin{align}
    \min_y \text{CVaR}_\alpha\big(\bm{X}(y)\big) = \min_{\bm{X}(y), z} \E{\max_{j\le J}l_{j}\big(\bm{X}(y)\big)}.\label{CVaROptShortcut}
\end{align}
Provided the expectation over $\bm{X}(y)$ is convex, then \eqref{CVaROptShortcut} is a simple convex optimization problem easily solved by many off-the-shelf solvers.

\subsection{Co-Control of VaR and CVaR in Stochastic Optimization}\label{subsec:CoControl}
Now, consider a general stochastic optimization problem of the following form
\begin{subequations}
\begin{align}
    \min_y \quad 
        & \mathbb{F}_0\{\bm{f}_0(y,\bm{X}_0)\}\label{StochObj}\\
    \text{s.t.} \quad 
        & \mathbb{F}_i\{\bm{f}_i(y,\bm{X}_i)\} \le b_i\quad i = 1,\hdots,M\label{StochCon}
\end{align}\label{StochGen}%
\end{subequations}%
for some mapping, $\mathbb{F}_i, i = 0,1,
\hdots, M$, of an uncertain quantity to a real number, which may differ for the evaluation of the objective in \eqref{StochObj} and each constraint in \eqref{StochCon}.
The choice of $\mathbb{F}_i$ will affect the solution to \eqref{StochGen}, and will depend on the needs of the decision maker.
Note here by affixing the subscript $i$ on each random vector $\bm{X}_i$ we allow for the possibility that the uncertainty in each term may arise from distinct random vectors not necessarily related to one another.
To allow for more general results, $\bm{X}_i$ denotes the random vector present in the $i$th term.
For instance in an optimal power flow (OPF) problem, for example, $\mathbb{F}_0$ may be the expectancy operator (i.e. $\mathbb{F}_0 \equiv \mathbb{E}$), and $\bm{X}_0$ some uncertain cost such that the objective is to minimize the expected cost, whereas $\mathbb{F}_i$ is commonly some $\VaR_\alpha$ or chance constraint that aims to maintain uncertain physical states $\bm{X}_i$ within their feasible limits.

Choosing $\mathbb{F}_i$ as a chance constraint or the $\VaR_\alpha$ may be difficult to work with in practical solutions as only a relatively few number of distributions offer exact deterministic reformulations, and the formulation for most distributions destroys convexity that may otherwise be available, \cite{rockafellar2007coherent}.
Furthermore, merely imposing a bound on the $\VaR_\alpha$ only provides a probabilistic guarantee on $\bm{f}_i$ not exceeding a certain value, with no measure of the expected loss if $\bm{f}_i$ exceeds that bound.
Although choosing $\mathbb{F}_i = \CVaR_\alpha$ provides both a probabilistic bound through the associated $\VaR_\alpha$ \textit{as well as} a sense of the risk of potential constraint violations, the CVaR on its own has some limitations.

If the $\CVaR_{\alpha}(\bm{f}_i(y,\bm{X}_i))$ $\le b_i$, then because $\CVaR_\alpha \ge \VaR_\alpha$ this implies that $\Pbig{\bm{f}_i(y,\bm{X}_i) \le b_i} \ge \alpha$ also holds as per  \eqref{VaRDef}.
Consequently, if we only want \eqref{StochCon} to hold with a probability $\alpha$ (i.e. a chance constraint on \eqref{StochCon}), then replacing this with a CVaR constraint such that $\CVaR_{\alpha}(\bm{f}_i(y,\bm{X}_i))$ $\le b_i$ will actually \textit{over-ensure} that the original constraint holds.
Hence, there may be a solution for some $\beta \le \alpha$ at which $\CVaR_{\beta}(\bm{f}_i(y,\bm{X}_i))\le b_i$ leads to $\VaR_\alpha(\bm{f}_i(y,\bm{X}_i)) = b_i$ such that we can relax the CVaR constraint and still ensure that the chance constraint holds, which may reduce the conservatism of the solution.

Finding $\beta$ such that $\beta < \alpha$ is not trivial, since it is not clear how $\alpha$ and $\beta$ are related, which this paper addresses.
Ultimately, we do not want the resulting VaR that comes from our calculation of the CVaR to be too conservative on our chance constraints, because it will affect our objective value.
As such, we want to not only minimize this CVaR at some level (i.e. control the CVaR), but since the $z_i$ defined in \eqref{CVaRMinimization} in that minimization is the VaR at that level, we also want to keep $z_i$ \textit{as close as possible} to critical value $b_i$ while ensuring we will not exceed it (i.e. co-control the VaR).
In order to accomplish this, while minimizing the CVaR of the constraints in \eqref{StochCon}, we would also like to constrain the VaR of each by penalizing deviations from upper limit $b_i$ without exceeding it.
This will control the VaR while simultaneously minimizing over the CVaR, i.e. co-controlling both quantities.

Thus we can choose measures $\mathbb{F}_0$ and $\mathbb{F}_i$ in \eqref{StochGen} to incorporate the CVaR of each term as in the optimization problem of \cite[Eq. (5.14)]{rockafellar2007coherent} and modify it to contain an additional constraint on $z_i$ of each chance constraint as follows:
\begin{subequations}
\begin{align}
    \min_{y,z,\rho \ge 0} \quad & \sum_{i = 0}^M\CVaR_{\alpha_i}(\bm{f}_i(y,\bm{X}_i),z_i) + \sum_{i = 1}^M \eta_i\rho_i\label{ModifiedObj}\\
    \text{s.t.} \quad
    &z_i + \rho_i = b_i \qquad \forall i\label{ModifiedConstraints}
\end{align}\label{CoControlProblemBase}%
\end{subequations}%
where $z \coloneqq \{z_0, z_1, \hdots, z_M\}$ is a vector of auxiliary decision variables necessary for the calculation of CVaR, whose value at the minimal of \eqref{CoControlProblemBase} is the VaR \cite{rockafellar2002conditional,rockafellar2007coherent}.
Furthermore, $\rho \coloneqq \{\rho_1,\rho_2,\hdots,\rho_M\}$ is a vector of positive slack variables penalized by parameters $\eta_i$ to ensure that the resulting VaR of each constraint is \textit{as close as possible} to the critical value $b_i$ without being greater than that.
Due to the relationship between the VaR and CVaR and the formulation of the problem, we have the following theorem:
\begin{theorem}\label{thm:rho}
The optimal $\rho_i^*$ in \eqref{CoControlProblemBase} is the exact amount by which the upper limit, $b_i$ exceeds the optimal VaR$_\alpha$.
\end{theorem}
\myproofstart
This comes from direct inspection of \eqref{ModifiedConstraints} and that $\rho_i \ge 0$, noting that these constitute a reformulation of the inequality $z_i \le b_i$ with positive slack variable $\rho_i$, and that $z_i$ is the VaR$_{\alpha_i}$ of each $\bm{f}_i$ from \eqref{SpecialFuncReform}.
\myproofend

Furthermore, with Theorem \ref{thm:rho} in place, for $\bm{f}_i$ with continuous, strictly increasing probability measures, which occurs for any $\bm{f}_i$ that has no probability masses like the normal or exponentially distributed random variable, we show:
\begin{corollary}
For a problem of the form of \eqref{CoControlProblemBase}, the optimal $\rho_i^*$ can be used to find the exact amount of added security for constraint $\bm{f}_i$ beyond the $\alpha_i$ tolerance level at the optimum, i.e. the amount by which $\Prob{\bm{f}^*_i \le b_i} > \alpha_i$.
\label{remark:addedSec}
\end{corollary}
\myproofstart
In the minimization of CVaR$_\alpha$, as defined in \cite{rockafellar2007coherent} and stated in \eqref{SpecialFuncReform}, $z_i$ is the resulting VaR$_\alpha$, which from \eqref{VaRDef} implies $\Prob{\bm{f}_i(y,\bm{X}_i) \le z_i} \ge \alpha_i$.
For $\bm{f}_i$ with continuous, strictly increasing probability measures, the VaR$_\alpha$ is the unique value, $z_i$ for which, \cite{rockafellar2002conditional}:
\begin{equation}
\Prob{\bm{f_i}(y,\bm{X}_i) \le z_i} = \alpha_i.\label{UniqueVaR}
\end{equation}
Note that $\Prob{\bm{f}_i \le b_i} \equiv \Prob{\bm{f}_i \le z_i + \rho_i}$ $ = \Prob{\bm{f}_i \le z_i} + \Prob{z_i \le \bm{f}_i \le z_i + \rho_i}$.
Substituting in \eqref{UniqueVaR}, we have that
\begin{equation}
    \Prob{\bm{f}_i \le b_i} = \alpha_i + \Prob{z_i \le \bm{f}_i \le z_i + \rho_i}\label{ExtraProb}
\end{equation}
i.e. solving \eqref{CoControlProblemBase} ensures $\bm{f}_i \le b_i$ with probability $\alpha_i + \Prob{z_i \le \bm{f}_i \le z_i + \rho_i}$, and $\Prob{z_i \le \bm{f}_i \le z_i + \rho_i}$ represents the exact amount of added security in the system.
\myproofend

We remark that, if the underlying distribution of $\bm{f}_i$ is known exactly, then one may correspondingly exactly calculate how much added security is in the system, and in cases where $\mathbb{P}$ is not known for $\bm{f}_i$, this quantity provides an estimate of added security.

\subsection{Wasserstein Metric in Data-Driven Stochastic Optimization}\label{sec:Wass}
Calculating \eqref{ModifiedObj} requires solving expectations with respect to a certain probability distribution of the $N-$dimensional random vector $\bm{X}_i$, see \eqref{SpecialFuncReform}.
If the form of the distribution is known, then there may exist exact reformulations of the expectations (e.g for a normally distributed variable), however generally we do \textit{not} know the true distribution and can only infer it from a finite set of historical samples.
Consequently, solutions of \eqref{CoControlProblemBase} that assume a specific distribution based on the available observation may exhibit poor out-of-sample performance.
With this in mind, we seek to employ a distributionally robust solution to \eqref{CoControlProblemBase} that accounts for all distributions that could have generated the available observations to further immunize against uncertainty.
Specifically, for a set of such candidate distributions (ambiguity set) $\mathcal{P}$, we minimize over the worst case distribution such that
\begin{subequations}
\begin{align}
    \min_{\substack{y,z,\\\rho \ge 0}} \quad 
        & \sup_{\mathbb{Q}\in \mathcal{P}}\quad \sum_{i = 0}^M\CVaR^{\mathbb{Q}}_{\alpha_i}(\bm{f}_i(y,\bm{X}_i),z_i) + \sum_{i = 1}^M \eta_i\rho_i \label{ModifiedObjectiveReform}\\
    \text{s.t.} \quad
        &  z_i + \rho_i = b_i \quad \forall i
\end{align}\label{droformulation}%
\end{subequations}%
where $\CVaR^{\mathbb{Q}}_{\alpha_i}(\bm{X})$ denotes the $\CVaR_{\alpha_i}(\bm{X})$ with respect to probability measure $\mathbb{Q}$ of $\bm{X}$.
Without proper construction of the ambiguity set, however, there is the possibility that the ambiguity set does not contain the true distribution, and inadequately accounts for uncertainty, or that the ambiguity set is too large, containing distributions that are not representative of the true underlying distribution, and thus yield an overly conservative solution.
Ideally we would like an ambiguity set that is just big enough to ensure that the true distribution is contained within, but not too big as to also contain many excess distributions.
To accomplish this, recent works have shown that a data-driven paradigm of constructing the ambiguity set using the Wasserstein metric leads to distributionally robust solutions that perform better than single distribution problems.
Furthermore, these solutions guarantee that the chance constraints with respect to the underlying true probability distribution can be robustly guaranteed based on a limited number of historical data points \cite{duan2018distributionally,esfahani2018data,gao2016distributionally}.
For the set, $\mathcal{P}$, the Wasserstein metric for any two distributions $\mathbb{Q}_1$, $\mathbb{Q}_2 \in \mathcal{P}$ can be defined as, \cite{esfahani2018data}:
\begin{align}
    W(\mathbb{Q}_1, \mathbb{Q}_2) \coloneqq \inf_\Pi \int_{\Omega^2}\norm{\bm{\omega}_1 - \bm{\omega}_2}\Pi(d\bm{\omega}_1,d\bm{\omega}_2)\label{WassersteinMetric}
\end{align}
where $\Pi$ is a joint distribution of $\bm{\omega}_1$ and $\bm{\omega}_2$ with marginal distributions $\mathbb{Q}_1$ and $\mathbb{Q}_2$, support $\Omega$, and $\norm{\cdot}$ can be any norm.

In order to construct an optimal ambiguity set with the Wasserstein metric, note that given the $K$-dimensional historical sample set $\{\hat{\omega}_1, \hat{\omega}_2,\hdots,\hat{\omega}_K\}$, the best estimate of the true distribution is empirical distribution $\hat{\mathbb{P}}_K(t) = \frac{1}{K}\sum_{i = 1}^K \bm{1}_{\{\hat{\omega}_i \le t\}},$ where $\bm{1}_{\{\hat{\omega}_i \le t\}}$ is the indicator function.
Furthermore, \cite{bolley2007quantitative, fournier2015rate} have shown that the unknown data-generating distribution belongs to the Wasserstein ambiguity set centered around $\EmpDist$ with confidence $1 - \beta$ if the corresponding Wasserstein radius grows sublinearly as a function of $\log(1/\beta)/K$.

Thus for empirical distribution $\EmpDist$ and true distribution $\mathbb{P}$ we have $W(\EmpDist, \mathbb{P}) \le \varepsilon(K)$ for some sample-dependent monotone function $\varepsilon(\cdot)$ that decreases to zero as its argument tends to infinity.
Accordingly, for a historical data set with $K$ samples, the true distribution $\mathbb{P}$ lies within the Wasserstein ball of radius $\epsilon(K)$ centered at empirical distribution $\EmpDist$ such that data-driven ambiguity set $\AmbSet$ is given by
\begin{align}
    \AmbSet = \big\{ \mathbb{P} \in \mathcal{P}(\Omega) | W(\mathbb{P},\EmpDist) \le \varepsilon(K)\big\}\label{eq:AmbiguitySet}
\end{align}
and represents the reliable information about the true distribution $\mathbb{P}$ observed from the $K$ historical samples.
To solve \eqref{droformulation}, we follow \cite{esfahani2018data} and set $\mathcal{P} = \hat{\mathcal{P}}_K$ as given by \eqref{eq:AmbiguitySet}.



Considering \eqref{droformulation} and using the definition of CVaR$_\alpha$ in \eqref{SpecialFuncReform} as the expectancy of the point wise maximum of loss function $l(\bm{f}(y,\bm{X}_i))$ along with \eqref{eq:AmbiguitySet} allows us to leverage the results of \cite{esfahani2018data} and state:

\vspace{-2mm}
\begin{theorem}
Let each $\bm{f}_i$ in \eqref{ModifiedObjectiveReform} be a convex function of both $y$ and $\bm{X}_i$, $\Omega_i \in \mathbb{R}^N$ be a closed, convex support of random vector $\bm{X}_i$, and $\AmbSet$ be the distribution set defined via the Wasserstein metric as in \eqref{eq:AmbiguitySet} with radius $\varepsilon(K) \ge 0$.
Defining the conjugate of a function, $[g(\cdot)]^* \coloneqq \sup_{\xi \in \mathbb{R}^N}\langle \cdot, \xi \rangle - g(\cdot)$, the dual norm, $||\cdot||_* \coloneqq \sup_{||\xi||\le 1}\langle \cdot, \xi \rangle$, and the support function of $\Omega_i$, $\sigma_{\Omega_i}(\cdot) \coloneqq \sup_{\omega \in \Omega_i}\langle \cdot, \omega \rangle$, and $l_{ji}(\bm{f}_i(y,\bm{X}_i)) \coloneqq a_{ji}\bm{f}_i(y,\bm{X}_i) + b_{ji}z_i$ for $a_{1i} = 0, b_{1i} = 1, a_{2i} = 1/(1 - \alpha_i), b_{2i} = 1 - 1/(1 - \alpha_i)$ as in Section \ref{sec:Prelim} for the CVaR$_\alpha$.
Then the problem in \eqref{droformulation} under the ambiguity set $\eqref{eq:AmbiguitySet}$ is equivalent to
\begin{subequations}
\begin{align}
    \min_{\substack{y,z,\rho \ge 0,\lambda, \\ s_{ik},t_{ikj},v_{ikj}}} & \quad \lambda \varepsilon(K) + \frac{1}{K}\sum_{k = 1}^K\sum_{i = 0}^ms_{ik} + \sum_{i = 1}^M \eta_i\rho_i \hspace{-3cm}&&\label{Thm6.1ReformObj}\\
    \text{s.t.}\qquad &  z_i + \rho_i = b_i && \forall i\\
    & [-l_{ji}]^*(t_{ikj} - v_{ikj}) + \sigma_{\Omega_i}(v_{ikj})\nonumber\\
    & \qquad - \langle t_{ikj}, \hat{\omega}_{ik} \rangle \le s_{ik} && \forall i,k,j\label{Thm6.1ReformCon1}\\
    & ||t_{ikj}||_* \le \lambda && \forall i,k,j\label{Thm6.1ReformCon2}
\end{align}\label{eq:Thm6.1Base}%
\end{subequations}%
where $\hat{\omega}_{ik}$ is the $k$th historical observation of random vector $i$, $\lambda,s_{ik},t_{ikj},v_{ikj}$ are auxiliary decision variables, $||\cdot||$ is a norm on $\mathbb{R}^N$ and $\langle a, b \rangle : = a^Tb$ denotes the inner product of two vectors, $a,b \in \mathbb{R}^N$.
\end{theorem}
\vspace{-2mm}
\myproofstart
We focus first on the CVaR term in objective \eqref{ModifiedObjectiveReform}, where we rewrite it using the formulation in \eqref{CVaROptShortcut}.
Namely, we have $\min_{y,z} \sup_{\mathbb{Q} \in \AmbSet} \E{\sum_{i = 0}^M \max_{j\le J} l_{ji}(\bm{f}_i(y,\bm{X}_i))}$.
By \cite[Theorem 6.1 and 4.2]{esfahani2018data}, for the Wasserstein ambiguity set this is equivalent to $\min_{y,z,\lambda,s_{ik},t_{ikj},v_{ikj}} \varepsilon(K) + \frac{1}{K}\sum_{k = 1}^K\sum_{i = 0}^ms_{ik}$ subject to constraints \eqref{Thm6.1ReformCon1} and \eqref{Thm6.1ReformCon2}.

The second term in \eqref{ModifiedObjectiveReform}, $\sum_{i = 1}^M \eta_i\rho_i$, is the sum of linear terms, and is therefore convex.
Thus adding this to the convex reformulation above preserves the overall convexity of the problem, and since the same term appears in \eqref{ModifiedObjectiveReform} and in the reformulation in \eqref{Thm6.1ReformObj}, the two problems remain equivalent.
Further, constraint \eqref{ModifiedConstraints} is affine, merely restricting the feasible set of values, and similarly for the convex constraint that $\rho_i \ge 0$, which restricts the feasible set to a half space.
Thus, since the feasible set of \eqref{droformulation} and \eqref{eq:Thm6.1Base} are equivalently restricted, the reformulation from \cite{esfahani2018data} remains valid, i.e. we have that \eqref{droformulation} with \eqref{eq:AmbiguitySet} is equivalent to \eqref{eq:Thm6.1Base}.
\myproofend

\vspace{-4mm}
While \eqref{eq:Thm6.1Base} holds for any convex loss functions that can be written as the point wise maximum of a piece-wise linear function, we may extend this in certain cases via the following corollary.
\vspace{-4mm}
\begin{corollary}
If the loss functions, $l_{ji}$, in \eqref{eq:Thm6.1Base} are affine functions, e.g. for loss functions of the form $l_{ji}(\bm{f}_i(y,\bm{X}_i) = a_{ji}\langle y, \bm{X}_i \rangle + b_{ji}z_i$, and the uncertainty sets are polytopes, i.e. $\Omega_i = \{ \bm{X}_i \in \mathbb{R}^N: C_i\bm{X}_i \le d_i\}$, where $C_i$ is a matrix and $d_i$ a vector of appropriate dimensions $\forall i$, then \eqref{eq:Thm6.1Base} must be smaller or equal to
\begin{subequations}
\begin{align}
    \min_{\substack{y,z,\rho \ge 0, \\ \lambda, s_{ik},\gamma_{ikj}}} & \quad \lambda \varepsilon(K) + \frac{1}{K}\sum_{k = 1}^K\sum_{i = 0}^Ms_{ik} + \sum_{i = 1}^M \eta_i\rho_i\hspace{-3cm}&&\label{Cor5.1GenReformObj}\\
    \text{s.t.}\qquad & z_i + \rho_i = b_i && \forall i\\
    & b_{ji}z_i + \langle \gamma_{ikj}, d_i - C_i\hat{\omega}_{ik}\rangle \nonumber\\
    & \qquad\qquad  + a_{ji}\langle y, \hat{\omega}_{ik}\rangle \le s_{ik} && \forall i,k,j\label{Cor5.1GenReformCon1}\\
    & ||C_i^T\gamma_{ikj} - a_{ji}y||_* \le \lambda && \forall i,k,j\label{Cor5.1GenReformCon2}\\
    & \gamma_{ikj} \ge 0 &&\forall i,k,j\label{Cor5.1GenReformCon3}
\end{align}\label{Cor5.1Base}
\end{subequations}
where the auxiliary decision variables are $\lambda,s_{ik},\gamma_{ikj}$
\end{corollary}
\vspace{-2mm}
\myproofstart
This follows directly from \cite[Corollary 5.1]{esfahani2018data} reformulating constraints \eqref{Thm6.1ReformCon1} and \eqref{Thm6.1ReformCon2} into \eqref{Cor5.1GenReformCon1} -- \eqref{Cor5.1GenReformCon3}.
\myproofend

\vspace{-5mm}
Naturally then, $\varepsilon(K)$ is crucial to the quality and conservatism of the result of \eqref{Cor5.1Base}.
As shown in \cite{duan2018distributionally}, following the derivations in \cite{zhao2015data,zhao2018data}, the radius can be expressed as
\begin{align}
    \varepsilon(K) = C\sqrt{\frac{1}{K}\log\bigg(\frac{1}{1 - \beta}\bigg)}\label{WassersteinRadius}
\end{align}
for some confidence level $\beta$ such that we may control the radius to be optimal size the data available via the value of $C$.
Further shown in \cite{duan2018distributionally}, we may safely estimate $C$ as
\begin{align}
    C \approx 2\inf_{\xi > 0} \Bigg(\frac{1}{2\xi}\bigg(1 + \ln\Big(\frac{1}{K}\sum_{i = 1}^Ke^{\xi\norm{\hat{\omega}_i - \hat{\mu}}_1^2}\Big)\bigg)\Bigg)^{1/2}\label{ApproxC}
\end{align}
where $\hat{\mu}$ is the sample mean of the data, and the minimization over $\xi$ can be performed by the bisection search method.

\section{Stochastic Reverse Auction}\label{sec:RA}

\begin{figure}[b]
    \centering
    \includegraphics[width=\columnwidth]{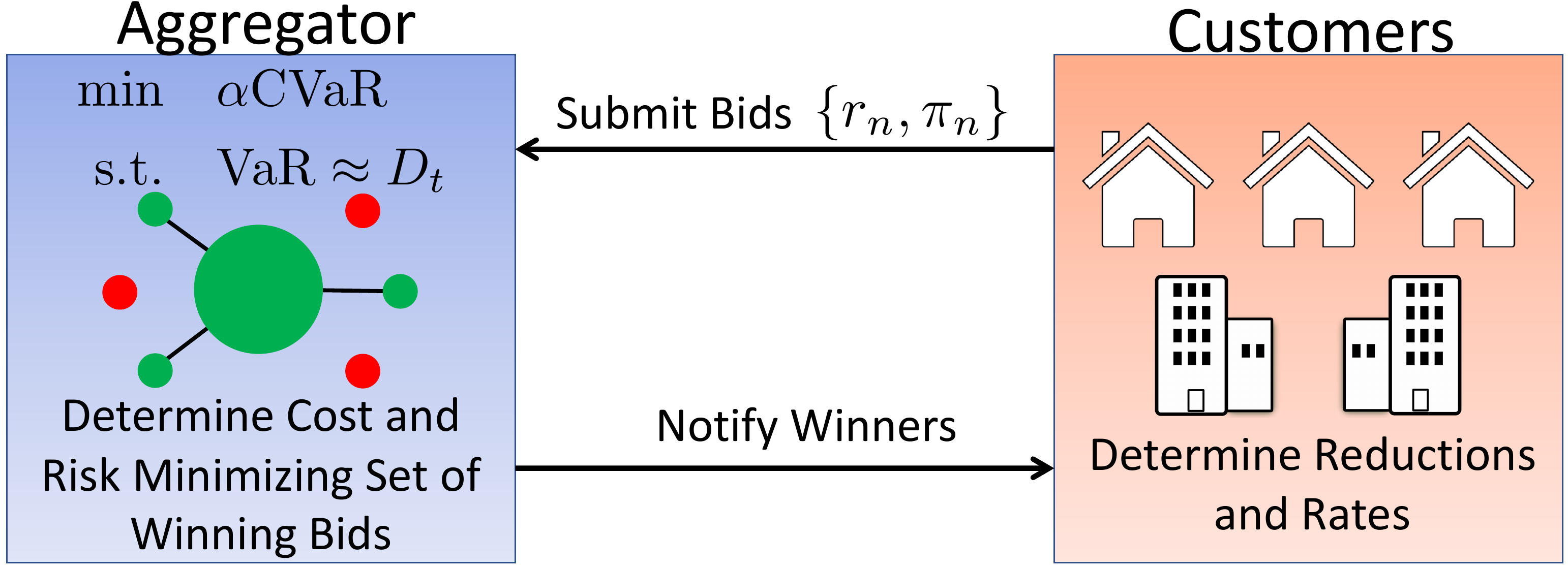}
    \vspace{-7mm}
    \caption{A schematic representation of the reverse auction. \label{fig:RA}}
    \vspace{-7mm}
\end{figure}

We turn now to a specific application of the above in which an aggregator operates a reverse auction for demand response (DR) as depicted in in Fig.~\ref{fig:RA}.

\subsection{Basic Problem}\label{subsec:RABasic}
The aggregator looks to receive a certain amount of DR $D_t$ determined by the power system operator, as in \cite{khezeli2017risk,li2017distributed}, and receives a penalty proportional to the amount by which they fail to meet $D_t$.
The aggregator procures DR from a number of customers $N$ within a DR market via a reverse auction.
Customers submit bids to the aggregator consisting of the amount they are willing to reduce their consumption from their base level before the call for DR $r_n$ as well as a price $\pi_n$ for each unit of reduced energy.

Furthermore, while customer's submit bids of how much they are willing to reduce their consumption, they may not need to reduce their consumption to this full amount, and are paid based on a realized reduction.
Because of either inaccuracy in the initial forecast of customers' consumption levels before the DR call or inability to actually deliver the scheduled amount of load consumption (voluntary or otherwise), the actual amount of delivered demand response of each customer is itself uncertain, rendering this a stochastic optimization problem.
Appropriately, we will model the actual reduction of each customer as the submitted bid $r_n$ plus an additional term to represent their unknown deviation from the bid $\bm{\Delta}_n$ which can be positive (over reduction), or negative (under reduction).
We assume that the aggregator will also have access to some historical record of each customer's bidding and \textit{actual} past consumption levels for the $k$th DR call $\check{x}_n^k$, i.e. the true amount of energy consumed at the time of past DR calls, as well as to the amount each customer has been \textit{scheduled} to consume $\hat{x}_n^k$.
Accordingly, the aggregator will be able to calculate a performance measure for each single DR call as $\delta_n^k \coloneqq \hat{x}_n^k - \check{x}_n^k$ for each customer $i$, which can be viewed as a \textit{realization} of the random variable $\bm{\Delta}_n$.

Hence, the loss function of the aggregator is the sum of all realized reduction levels $\bm{\xi}_n \coloneqq r_n + \bm{\Delta}_n$ multiplied by price $\pi_n$ from all \textit{accepted} bids.
For ease of notation we define $\bm{\omega}_0 \coloneqq \{\bm{\xi}_1\pi_1,\hdots, \bm{\xi}_N\pi_N\}$ to be the $N$-dimensional vector of uncertain customer reductions multiplied by the price at which they wish to be compensated such that $\bm{\omega}_{0n} = \bm{\xi}_{n}\pi_n$, the uncertain reduction of customer $n$ multiplied by customer $n$'s desired compensation, 
Similarly, we define $\bm{\omega}_1 \coloneqq \{\bm{\xi}_1,\hdots, \bm{\xi}_N\}$ to be the $N$-dimensional vector of uncertain customer reductions.
The aggregator therefore tries to optimally select the bids minimize this loss function.
However, the aggregator must also ensure that the sum of all actual reduction levels (i.e. the DR deliverable to the system operator) is greater than their target goal, $D_t$, i.e. for the vector of binary decisions $u \coloneqq \{u_1, u_2, \hdots, u_N\}$
\begin{subequations}
\begin{align}
    \min_{u} \quad & \langle u, \bm{\omega}_0 \rangle\label{StochWSPObj}\\
    \text{s. t.} \quad & -\langle u, \bm{\omega}_1\rangle \le -D_t\label{MinDR}\\
    & u_n \in \{0,1\}\label{BinaryDecision}
\end{align}\label{RABase}%
\end{subequations}%
\vspace{-4mm}

\subsection{Co-Controlled Distributionally Robust Reformulation}\label{subsec:RAReform}
This section recasts \eqref{RABase} in the context of Section~\ref{sec:CoControl}, i.e. to ensure \eqref{MinDR} holds with a given probability while minimizing the harmful tail costs of both \eqref{StochWSPObj} and \eqref{MinDR} (i.e. minimizing the $\CVaR_\alpha$ of both) in a data-driven distributionally robust manner.
Before proceeding, note that by affixing the superscript $k$ and adding a $\hat{\cdot}$ to the vectors $\bm{\omega}_0$ and $\bm{\omega}_1$, we define $\hat{\omega}_0^k$ as the the $k$th observation of the vector $\bm{\omega}_0$, and similarly for for $\hat{\omega}_1^k$.
As such, $\hat{\omega}_0^k$ and $\hat{\omega}_1^k$ are no longer uncertain, but are realizations of the random variables.
Now, we reformulate the problem from \eqref{RABase} with co-control CVaR and VaR considerations in a distributionally robust manner using the ambiguity set defined via the Wasserstein metric and $K$ historical observations of each customer using our results from Sections \ref{subsec:CoControl} and \ref{sec:Wass} as follows:
\begin{subequations}
\begin{align}
    \min_{\substack{u, z,\rho \\ \lambda,s_{ik},\gamma_{ikj}}} & \quad \lambda \varepsilon(K) + \frac{1}{K}\sum_{k = 1}^K\sum_{i = 0}^1s_{ik} + \eta\rho\label{Cor5.1ReformObj}\\
    \text{s.t.}\qquad & \text{\eqref{Cor5.1GenReformCon1} -- \eqref{Cor5.1GenReformCon3}}\\
    & z_1 + \rho = -D_t\label{CoControlMinDR}\\
    & u_n \in \{0,1\}
\end{align}\label{RAReformBase}%
\end{subequations}%
Similar to Section ~\ref{sec:Wass}, $J = 2$, $a_{1i} = 0, b_{1i} = 1, b_{2i} = 1 - 1/(1 - \alpha_i)$ $i \in \{0,1\}, a_{20} = 1/(1 - \alpha_i), a_{21} = -1/(1 - \alpha_i)$ , all $\alpha_i$ are the tolerance to risk, i.e. the probability with which one desires a certain constraint to hold.
Additionally all $z_i$ are the VaR, controlled for the constraint given by \eqref{MinDR} via \eqref{CoControlMinDR}, $\eta$ is some number characterizing the decision maker's preference of added security in the system, and $\rho$ is a positive slack variable that allows for control.
Thus solving \eqref{RAReformBase} allows us to find the least cost set of winners of the DR reverse auction that ensures the target DR is met, while simultaneously controlling the VaR and CVaR in a data-driven distributionally robust manner.

\section{Numerical Results}

This case study considers a stochastic reverse auction  in Fig.~\ref{fig:RA} as modeled  in  \eqref{RAReformBase} with 10 customers using the Pecan Street data \cite{street2015dataport} from the New York area. Each customer is allowed  to bid up to $\gamma$ of their average energy consumption, for $\gamma$ ranging from 10\% to 30\%, at the unitary price $\pi_n=1$.  The penalty factor $\eta$ is set as a fraction $\eta^{\pi}$ of $\eta$ and the total amount of DR to be collected ($D_t$) is set to 50\% of the collected bids.  To solve \eqref{RAReformBase},  we generated 100 random samples in $\Omega_i$ with the variance of samples $p$ of the bid,  with $p$ varying from 10\% to 30\%. In other words, customers with a smaller bid had a smaller variance, and customers with a larger bid had a larger variance, which allows for studying a trade-off between low-risk, low-return customers (i.e. small bid, small variance) and higher-risk, higher-return customers (i.e. those with a larger bid but also a larger variance).
All simulations were conducted using Julia JuMP package and the Gurobi solver \cite{DunningHuchetteLubin2017, gurobi}.

Table~\ref{tab:rhoresults} compares the performance of \eqref{RAReformBase} for different penalty values. We use the case with no penalty ($\eta^{\pi}=0$), i.e. CVaR is minimized with no co-control on VaR, to establish the base case. As $\eta^{\pi}$ increases, so does the percent decrease in $\rho$ for all combinations of $\gamma$ and $\sigma$. In other words, penalty $\eta$ reduces a  gap between the VaR and a desired target goal $D_t$. Notably, while Table ~\ref{tab:rhoresults} demonstrates that co-controlling enabling constraint \eqref{CoControlMinDR} reduces the gap between the VaR and target goal, the size of the gap depends on the variance and a larger variance  yields a lower gap reduction. 
\begin{table}[h!] 
\caption{Percentage Differences of $\rho$ for Various $\eta$, $\gamma$, and $\sigma$}
\label{tab:rhoresults}
\setlength{\tabcolsep}{5pt}
  \begin{center}
    \vspace{-2mm}
    \begin{tabular}{|c|c|c|c|c|c|c|c|c|c|}
        \hline
        & \multicolumn{3}{c|}{$\gamma = 10\%$} & \multicolumn{3}{|c|}{$\gamma = 20\%$} & \multicolumn{3}{|c|}{$\gamma = 30\%$} \\
        \hline
        \backslashbox{$\sigma$}{$\eta^{\pi}$} & 0.0 & 0.5 & 1.0 & 0.0 & 0.5 & 1.0 & 0.0 & 0.5 & 1.0\\
        \hline
        $10\%$ & 0.0 & 99.8 & 100 & 0.0 & 99.4 & 100 & 0.0 & 29.7 & 100\\
        \hline
        $20\%$ & 0.0 & 99.9 & 100 & 0.0 & 98.2 & 100 & 0.0 & 59.9 & 100\\
        \hline
        $30\%$ & 0.0 & 86.5 & 100 & 0.0 & 93.7 & 100 & 0.0 & 73.0 & 100\\
        \hline
    \end{tabular}
  \end{center}
  \vspace{-3mm}
\end{table}

\vspace{-2mm}
\section{Conclusion}
This paper has presented a method to calculate the CVaR of a constraint from a finite number of samples, while simultaneously controlling the VaR level to be as close as possible without exceeding the upper bound of a constraint. This co-control method has also been extended to  accommodate the underlying uncertainty in a distributionally robust manner. Building on the existing result in \cite{esfahani2018data}, we proved that the CVaR and VaR co-control over an ambiguous uncertainty set can be implemented in a computationally tractable manner.  The proposed CVaR and VaR co-control  has been applied to a stochastic reverse auction,  in which an aggregator seeks to procure uncertain amounts of DR from a pool of customers at a lowest cost, similar to a portfolio optimization problem.
Our results demonstrate that the proposed control scheme effectively reduces the gap between the VaR and the upper bound of a constraint.
Future efforts will investigate the effects of this co-control model on revenue adequacy and cost recovery properties of the reverse auction in the context of distribution power flow  constraints

\bibliographystyle{IEEEtran}
\bibliography{bib.bib}

\begin{thebibliography}{10}
\providecommand{\url}[1]{#1}
\csname url@samestyle\endcsname
\providecommand{\newblock}{\relax}
\providecommand{\bibinfo}[2]{#2}
\providecommand{\BIBentrySTDinterwordspacing}{\spaceskip=0pt\relax}
\providecommand{\BIBentryALTinterwordstretchfactor}{4}
\providecommand{\BIBentryALTinterwordspacing}{\spaceskip=\fontdimen2\font plus
\BIBentryALTinterwordstretchfactor\fontdimen3\font minus
  \fontdimen4\font\relax}
\providecommand{\BIBforeignlanguage}[2]{{%
\expandafter\ifx\csname l@#1\endcsname\relax
\typeout{** WARNING: IEEEtran.bst: No hyphenation pattern has been}%
\typeout{** loaded for the language `#1'. Using the pattern for}%
\typeout{** the default language instead.}%
\else
\language=\csname l@#1\endcsname
\fi
#2}}
\providecommand{\BIBdecl}{\relax}
\BIBdecl

\bibitem{pritchard2010single}
G.~Pritchard, G.~Zakeri, and A.~Philpott, ``A single-settlement, energy-only
  electric power market for unpredictable and intermittent participants,''
  \emph{Operations research}, vol.~58, no. 4-part-2, pp. 1210--1219, 2010.

\bibitem{papavasiliou2011reserve}
A.~Papavasiliou, S.~S. Oren, and R.~P. O'Neill, ``Reserve requirements for wind
  power integration: A scenario-based stochastic programming framework,''
  \emph{IEEE Tran. Pwr. Syst.}, vol.~26, no.~4, pp. 2197--2206, 2011.

\bibitem{zheng2013decomposition}
Q.~P. Zheng, J.~Wang, P.~M. Pardalos, and Y.~Guan, ``A decomposition approach
  to the two-stage stochastic unit commitment problem,'' \emph{Annals of
  Operations Research}, vol. 210, no.~1, pp. 387--410, 2013.

\bibitem{bienstock2014chance}
D.~Bienstock, M.~Chertkov, and S.~Harnett, ``Chance-constrained optimal power
  flow: Risk-aware network control under uncertainty,'' \emph{Siam Review},
  vol.~56, no.~3, pp. 461--495, 2014.

\bibitem{ben2009robust}
A.~Ben-Tal, L.~El~Ghaoui, and A.~Nemirovski, \emph{Robust optimization}.\hskip
  1em plus 0.5em minus 0.4em\relax Princeton University Press, 2009, vol.~28.

\bibitem{lubin2015robust}
M.~Lubin, Y.~Dvorkin, and S.~Backhaus, ``A robust approach to chance
  constrained optimal power flow with renewable generation,'' \emph{IEEE Tran.
  Pwr. Syst.}, vol.~31, no.~5, pp. 3840--3849, 2015.

\bibitem{zhao2018data}
C.~Zhao and Y.~Guan, ``Data-driven risk-averse stochastic optimization with
  wasserstein metric,'' \emph{Op. Res. Let.}, vol.~46, no.~2, pp. 262--7, 2018.

\bibitem{delage2010distributionally}
E.~Delage and Y.~Ye, ``Distributionally robust optimization under moment
  uncertainty with application to data-driven problems,'' \emph{Operations
  research}, vol.~58, no.~3, pp. 595--612, 2010.

\bibitem{erdougan2006ambiguous}
E.~Erdo{\u{g}}an and G.~Iyengar, ``Ambiguous chance constrained and robust
  optimization,'' \emph{Math. Prog.}, vol. 107, no. 1-2, pp. 37--61, 2006.

\bibitem{mieth2018data}
R.~Mieth and Y.~Dvorkin, ``Data-driven distributionally robust optimal power
  flow for distribution systems,'' \emph{IEEE Control Systems Letters}, vol.~2,
  no.~3, pp. 363--368, 2018.

\bibitem{esfahani2018data}
P.~Esfahani and D.~Kuhn, ``Data-driven distributionally robust optimization
  using the wasserstein metric: Performance guarantees and tractable
  reformulations,'' \emph{Math. Prog.}, vol. 171, no. 1-2, pp. 115--166, 2018.

\bibitem{lim2010portfolio}
C.~Lim, H.~Sherali, and S.~Uryasev, ``Portfolio optimization by minimizing
  conditional value-at-risk via nondifferentiable optimization,'' \emph{Comp.
  Opt. and App.}, vol.~46, no.~3, pp. 391--415, 2010.

\bibitem{rockafellar2007coherent}
R.~T. Rockafellar, ``Coherent approaches to risk in optimization under
  uncertainty,'' in \emph{OR Tools and App.: Gl. of Fut. Tech.}, 2007, pp.
  38--61.

\bibitem{rockafellar2000optimization}
R.~T. Rockafellar \emph{et~al.}, ``Optimization of conditional value-at-risk,''
  \emph{Journal of risk}, vol.~2, pp. 21--42, 2000.

\bibitem{rockafellar2002conditional}
R.~T. Rockafellar and S.~Uryasev, ``Conditional value-at-risk for general loss
  distributions,'' \emph{J. Bank. \& Fin.}, vol.~26, no.~7, pp. 1443--1471,
  2002.

\bibitem{dall2017chance}
E.~Dall’Anese, K.~Baker, and T.~Summers, ``Chance-constrained ac optimal
  power flow for distribution systems with renewables,'' \emph{IEEE Tran. Pwr.
  Syst.}, vol.~32, no.~5, pp. 3427--3438, 2017.

\bibitem{bolley2007quantitative}
F.~Bolley \emph{et~al.}, ``Quantitative concentration inequalities for
  empirical measures on non-compact spaces,'' \emph{Prob. Th.}, vol. 137, pp.
  541--93, 2007.

\bibitem{fournier2015rate}
N.~Fournier and A.~Guillin, ``On the rate of convergence in wasserstein
  distance of the empirical measure,'' \emph{Probability Theory and Related
  Fields}, vol. 162, no. 3-4, pp. 707--738, 2015.

\bibitem{duan2018distributionally}
C.~Duan, W.~Fang, L.~Jiang, L.~Yao, and J.~Liu, ``Distributionally robust
  chance-constrained approximate ac-opf with wasserstein metric,'' \emph{IEEE
  Tran. Pwr. Syst.}, vol.~33, no.~5, pp. 4924--4936, 2018.

\bibitem{gao2016distributionally}
R.~Gao and A.~J. Kleywegt, ``Distributionally robust stochastic optimization
  with wasserstein distance,'' \emph{arXiv preprint arXiv:1604.02199}, 2016.

\bibitem{zhao2015data}
C.~Zhao and Y.~Guan, ``Data-driven risk-averse stochastic program with
  $\zeta$-structure probability metrics,'' \emph{Available on Optimization
  Online}, 2015.

\bibitem{khezeli2017risk}
K.~Khezeli and E.~Bitar, ``Risk-sensitive learning and pricing for demand
  response,'' \emph{IEEE Tran. Sm. Gr.}, vol.~9, no.~6, pp. 6000--6007, 2017.

\bibitem{li2017distributed}
{P. Li et al}, ``A distributed online pricing strategy for demand response
  programs,'' \emph{IEEE Tran. Sm. Gr.}, vol.~10, no.~1, pp. 350--360, 2017.

\bibitem{street2015dataport}
P.~Street, ``Dataport: the world's largest energy data resource,'' \emph{Pecan
  Street Inc}, 2015.

\bibitem{DunningHuchetteLubin2017}
I.~Dunning, J.~Huchette, and M.~Lubin, ``Jump: A modeling language for
  mathematical optimization,'' \emph{SIAM Review}, vol.~59, no.~2, pp.
  295--320, 2017.

\bibitem{gurobi}
\BIBentryALTinterwordspacing
L.~Gurobi~Optimization, ``Gurobi optimizer reference manual,'' 2020. [Online].
  Available: \url{http://www.gurobi.com}
\BIBentrySTDinterwordspacing

\end{thebibliography}

\end{document}